\documentclass[%
 reprint,
 superscriptaddress,
 amsmath,amssymb,
 aps,
 prl,
]{revtex4-2}

\usepackage[utf8]{inputenc}
\usepackage[english]{babel}
\usepackage[T1]{fontenc}
\usepackage{color}
\usepackage[colorinlistoftodos,prependcaption,textsize=small]{todonotes}

\usepackage[export]{adjustbox}

\definecolor{strongpink}{HTML}{AA336A}

\usepackage{graphicx}
\usepackage{dcolumn}
\usepackage{bm}
\usepackage{url} 
\usepackage[colorlinks=true,citecolor=blue]{hyperref} 

\begin{document}

\preprint{APS/123-QED}

\title{Goldilocks fluctuations: dynamic constraints on loop formation\\
in scale-free transport networks}


\author{Radost Waszkiewicz}
\affiliation{
Institute of Theoretical Physics, Faculty of Physics, University of Warsaw, Pasteura 5, 02-093 Warsaw, Poland 
}

\author{John Burnham Shaw}
\affiliation{
Department of Geosciences, University of Arkansas, Fayetteville, AR, USA
}

\author{Maciej Lisicki}
\affiliation{
Institute of Theoretical Physics, Faculty of Physics, University of Warsaw, Pasteura 5, 02-093 Warsaw, Poland 
}

\author{Piotr Szymczak}
\affiliation{
Institute of Theoretical Physics, Faculty of Physics, University of Warsaw, Pasteura 5, 02-093 Warsaw, Poland 
}

\date{\today}

\begin{abstract}
Adaptive transport networks are known to contain loops when subject to hydrodynamic fluctuations. However, fluctuations are no guarantee that a loop will form, as shown by loop-free networks driven by oscillating flows. We provide a complete stability analysis of the dynamical behaviour of any loop formed by fluctuating flows. We find a threshold for loop stability that involves an interplay of geometric constraints and hydrodynamic forcing mapped to constant and fluctuating components. Loops require fluctuation in the relative size of the flux between nodes, not just a temporal variation in the flux at a given node. Hence, there is both a minimum and a maximum amount of fluctuation relative to the constant-flux component where loops are supported.
\end{abstract}

\maketitle

Transport networks are fundamental components of many complex systems, including blood vasculature~\cite{Schneider2012}, metabolic flux distribution networks~\cite{segre_analysis_2002},  river~\cite{smart_quantitative_1971-1}
or karst conduit systems~\cite{Palmer1991}. Understanding the mechanisms behind their formation and growth is crucial for finding a link between the form, function, and efficiency of the networks. Are spontaneously-formed networks optimal, e.g. in terms of minimal energy dissipation~\cite{Rinaldo1993,Maritan1996,Banavar_2000,Ronellenfitsch2016} or global resistance~\cite{Errera1998,Bejan2006}? If yes, then to what topologies do the optimal structures correspond? 
The answers to these questions remain elusive. Even more elusive is the link between the final geometries of the networks and their growth dynamics: it is not clear whether spontaneous growth of the network leads to optimal structures~\cite{Devauchelle2017}, although pruning of the less effective branches might help circumvent this problem~\cite{Ronellenfitsch2016}. 

\begin{figure*}
    \centering
    \includegraphics[width=\linewidth]{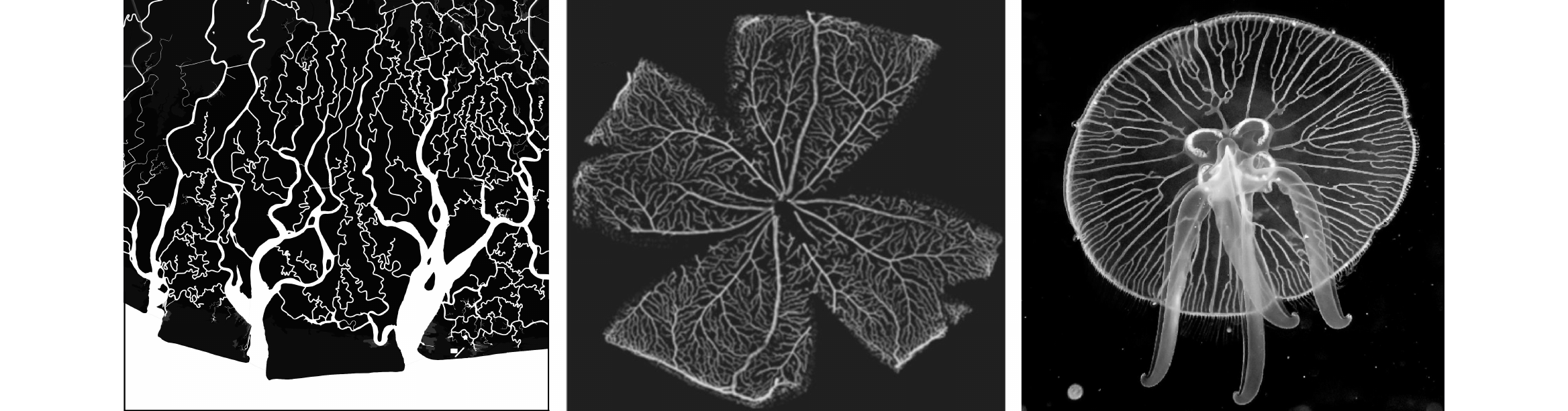}
    \caption{Examples of reticulated networks with different geometries and topologies. Left; a land-water binary map of the Niger river delta (copyright: OpenStreetMap contributors, data under ODbL, \cite{OpenStreetMap_Niger_Delta}) reveals a scale-free loopy network \cite{George_2019}. Center; vasculature in a mouse retina (copyright: Bernabeu et al, CC BY 3.0 \cite{Bernabeu_2014}), a loopy but scale-dependent network. Right; gastrovascular canal system of a jellyfish {\it Aurelia aurita} (copyright: OISTGU, CC BY 4.0 \cite{Moon_Jelly}) with only a few loops. Channels transporting fresh water and nutrients change during the growth of the animal with an interplay between hydrodynamic and mechanical stimuli.
    } 
    \label{fig:illustrations}
\end{figure*}

Network topologies range from highly ramified hierarchical trees to well-connected loopy structures~\cite{Banavar_2000,Durand2007,Ronellenfitsch2019}, but the reasons for such a variety are still debated. In has been suggested that the topology is controlled by the form of the cost function for local transportation of material, which is minimised in optimal structures \cite{Banavar_2000,Bohn_2007,Durand2007}. Alternately, loops could arise from a trade-off between cost and resilience to random damage incurred to the network~\cite{Katifori_2010,Kaiser_2020}: trees dominate in systems where cost minimisation is paramount, e.g. power grids feeding a small number of homes or large blood vessels distributing blood to the organs. In contrast, loops dominate when the cost of a single connection is relatively low, yet a system must remain resilient to the threat of damage~\cite{ganin_resilience_2017}, as observed in leaf venations~\cite{Mitchison1980} or capillary plexus~\cite{Campbell2017}. 
Overall, it is conceivable that there are multiple mechanisms of loop formation in the transport networks, differing in the resulting geometrical patterns. 
Fig. \ref{fig:illustrations} shows three examples of looping structures. The first is a scale-free, fractal loopy network of water channels in the Niger River. The second is a scale-dependent network of veinlets and arterioles in mouse retina \cite{Bernabeu_2014}, with a tree-like structure on larger scale and loopy capillary plexus on smaller scale~\cite{Ronellenfitsch2019}. The third photo shows a network of gastrovascular canals in an {\it Aurelia aurita} jellyfish, with a characteristic, hierarchical structure with loops~\cite{Song2023}. 

The growth dynamics have received less scrutiny compared to steady state of networks discussed above. 
It is not easy to obtain loops as a result of growth driven by the external field: most simple growth models, in which one phase grows at the expense of another, produce loopless structures. Examples include Laplacian growth phenomena~\cite{Gustafsson2014}, such as diffusion-limited aggregation~\cite{Witten1981}, dielectric breakdown~\cite{Niemeyer1984}, electrodeposition~\cite{Grier1986}, all resulting in
tree-like fractal structures. Loops do emerge if the finite mobility of the invading phase is accounted for, as shorter branches of the growing structures become attracted to the longer ones forming a nested loop structure~\cite{Budek2017}, resembling the gastrovascular canal patterns of jellyfish, shown in Fig. \ref{fig:illustrations}c.

The emergence of loopy topologies has also been linked to the growth of a network in the presence of fluctuations in the flux~\cite{Corson_2010,Katifori_2010,Hu2013}. If the characteristic timescales of fluctuations are shorter than the network adaptation times and their amplitudes are relatively large, loops tend to appear in the system. 
However, the emergent equilibria of many optimal networks are not entirely loopy: a significant fraction of the possible links have a negligible conductivity, and many subnetworks are trees. 
Hence, while the controls on the global ``loopiness'' of the network are beginning to be understood, the constraints on what makes a particular domain a loop remain enigmatic.

Recent efforts have explored the transition between topologies containing few and many loops in network geometries that minimize dissipation. Kaiser et al.~\cite{Kaiser_2020} considered a simple loop within a network, in which nodes that play the roles of sources and sinks are driven by independent fluctuations of the flux. They showed that beyond a threshold of flux variance, a loop became optimal relative to a tree. While helpful, this advance cannot describe the enigma of 
many natural networks, e.g. the channel networks in coastal salt marshes~\cite{rinaldo_tidal_1999-1}, which are controlled by strong tidal fluctuations and yet remain loopless. These networks are governed by tidal flows that reverse on a diurnal or semi-diurnal timescale, but where sources and sinks are highly correlated at short lengthscales~\cite{sullivan_complexity_2015,hale_observations_2019}, which remove loops as effectively as constant flows~\cite{Konkol_2022}. It is clear that aspects of the applied flows beyond the magnitude of fluctuations are important, but remain presently unexplained.

\begin{figure*}
    \centering

    \includegraphics[width=\linewidth]{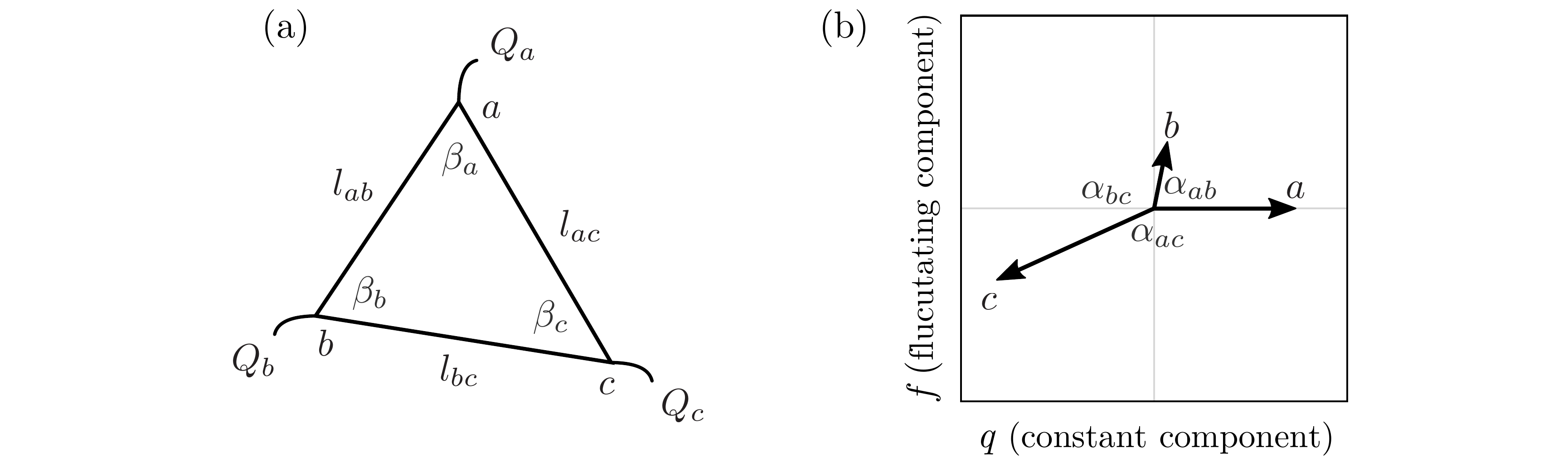}
    \caption{
    \textbf{a)} Sketch of a single triangular loop.     
    Three vertices subject to fluctuating discharges $Q_{i}$ are connected by channels of conductivity $k_{ij}$. Three angles $\beta_{i}$ characterise the geometry of the loop.
    \textbf{b)} Inflow characterisation in $q,f$ plane corresponding to the situation in panel (a). 
    Node $a$ is fed by a constant flow, whereas node $b$ is fed by a fluctuating source. Node $c$ is an outlet (with negative $q$ and $f$) with a flow which is a sum of that in $a$ and $b$ and hence representing a mixture of constant and fluctuating component. Angles $\alpha_{ij}$ between the flow vectors control the stability of the system. }
    \label{fig:sketch}
\end{figure*}

In this paper, we attempt to elucidate the local
factors controlling the formation of loops in
transport networks by analysing 
 loop stability in the simplest possible model system consisting of a single triangular loop, as shown in Fig.~\ref{fig:sketch}a.
In the case of a river delta, one can imagine the links as channels carrying water and nodes as locations where the channels meet. Because of fluctuations in the system driven by tides or river floods (or explicit water management), flows from the rest of the network to the nodes vary over time. 
We will treat the rest of the network as a source of (time-dependent) volumetric flux into the channels. Such a simplification is adequate when the size of the loop is small in comparison to the network in general, and changes to the conductivities of the links in the loop do not affect the forcing flows. 


The dynamics of a channel network are governed by relationships for flows and adaptation. For flows, we assume a linear relationship between pressure gradient and discharge $Q$ through a link, scaled by conductivity $k$ and link length $l$.
\begin{equation}
    Q(k) = k \frac{\Delta p}{l},\label{eqn:ohm_law}
\end{equation}
which holds for laminar flow in small vessels of cross-sectional area $A$, such as leaf veins~\citep{dimitrov_constant_2006,Katifori_2010} and blood capillaries~\citep{murray_physiological_1926,pries_structural_2001} where the Hagen-Poiseuille law applies and the conductivity $k$ can be expressed as $k = A^2 / 8\pi\eta$, where $\eta$ is the dynamic viscosity of the fluid.
This model is also an adequate approximation of friction-dominated shallow-water flows in coastal networks~\citep{Rinaldo_1999,Van_Oyen_2012,Van_Oyen_2014}. 

Next, we assume that the networks evolve sufficiently slowly, adapting the diameters of their links (and hence also conductivities) to the flows. Such adaptation is often assumed to be a process of optimization, in which the channel evolves to minimize power dissipation expressed as $Q^2/k$ (or equivalently $Q\Delta{p}/l$)~\citep{Banavar_2000,Rinaldo_2006,Bohn_2007,Katifori_2010,Corson_2010,Kaiser_2020,Konkol_2022}. This approach motivates the functional relationship between the rate of change of conductivity $k$ and the average squared discharge $Q^2$ and the conductivity itself. Empirical observations of such networks at equilibrium show a nonlinear relationship between $k$ and the average $Q^2$ (smoothing out fast fluctuations, such as daily fluctuations in tidal cycles) denoted by $\langle Q^2 \rangle$. In a stationary state, we typically approximate this relationship by a power law $k = a \tau \langle Q(k)^2 \rangle^\gamma$.
As a result, one can propose a dynamical equation~\cite{Tero2010,Hu2013,Konkol_2022}
\begin{equation}
    \frac{d}{dt} k(t) = a \langle Q(k)^2 \rangle ^\gamma - \frac{1}{\tau} k(t) \label{eqn:adapt_optimize}
\end{equation}
which reproduces such a power law at equilibrium. Here, $a$ and $\gamma$ control the shape of the power law, while $\tau$ controls the time scale of relaxation to equilibrium. For coastal rivers, empirical scaling suggests that $\gamma\approx 0.6$ \citep{Konkol_2022}. In this study, taking a slightly lower value of $\gamma=1/2$ allows us to find an analytical solution to a forthcoming linear stability analysis. By a suitable choice of units for $Q$ and $t$, we can cast Eq.~\eqref{eqn:adapt_optimize} in a dimensionless form in which $a=\tau=1$. We will assume such a choice of units in subsequent considerations. 

Each node of our triangular network is also connected to a flow source $Q_i$, which mimics interactions with the remaining links of a large network ({\it cf}. Fig.~\ref{fig:sketch}a). Labeling the vertices $a,b,c$ and the links $ab,bc,ca$ we can write the conservation of mass in the system as $Q_{ca} - Q_{ab} = Q_a$ and cyclically. 
Combining this with the pressure equation~\eqref{eqn:ohm_law} and solving for flows leads to
\begin{equation}
\begin{split}
    Q_{ab} &= \frac{C_{ab} \left(Q_b C_{ca} - Q_a C_{bc}\right) }{\sigma_2} \\
    \sigma_2 &= C_{ab}C_{bc} + C_{bc}C_{ca} + C_{ca}C_{ab} 
    \label{eqn:Qij}
\end{split}
\end{equation}
and cyclically. In the above, $C = k/l$ is the ratio of conductivity and length defined for each of the links. 

Let us now imagine that the influxes to our network fluctuate. Combining Eq.~\eqref{eqn:Qij} and 
Eq.~\eqref{eqn:adapt_optimize}, we can express the evolution of conductivity of each channel using averaged products of the forcing fluxes. More concretely, we define a dynamical system in the $(C_{ab},C_{bc},C_{ca})$ space (or equivalently, the $(k_{ab},k_{bc},k_{ca})$ space) given by

\begin{equation}
\frac{\mathrm{d}}{\mathrm{d}t} k_{ab} = \frac{C_{ab}}{\sigma_2} \langle (Q_b C_{ca} - Q_a C_{bc})^2 \rangle^{1/2} - k_{ab}
\label{eqn:dynamial_system}
\end{equation}
and cyclically. In principle, one can find an explicit solution for the fixed points of the above dynamical system, but the expression is lengthy and not particularly informative. A more insightful procedure is to assume that initially one of the links (e.g. $ab$) had zero conductivity. The immediate conclusion from Eq.~\eqref{eqn:Qij} is that then $C_{ab}$ will remain zero at all times. We can solve for flows in the two remaining links and compute their stationary conductivities to be
\begin{equation}
C_{ab}=0, \ \ C_{bc}=\sqrt{\langle Q_{b}^2 \rangle}, \ \ C_{ca}=\sqrt{\langle Q_a^2 \rangle}.
    \label{eqn:fixed_wall}
\end{equation}
The stability of the fixed point defined by Eq.~\eqref{eqn:fixed_wall} can then be established by linear stability analysis, where a stable solution means that a small positive $C_{ab}$ will shrink back to zero (maintaining a tree) while an unstable solution will grow and support a loop. 

\begin{figure*}
    \centering
    \includegraphics[width=\linewidth]{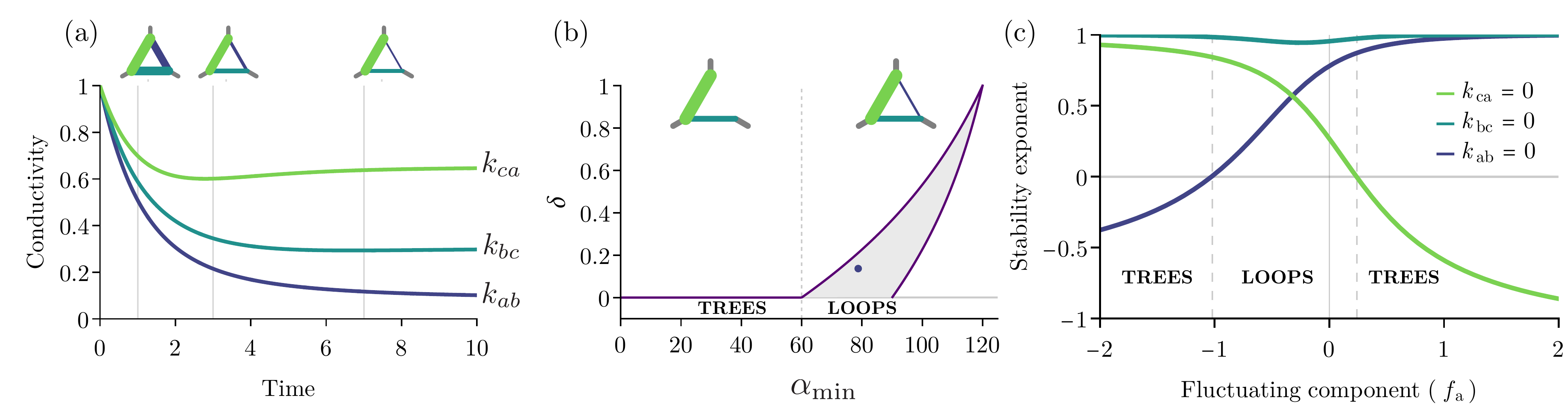}
    \caption{
    \textbf{a)} Evolution of the link conductivities of the triangular network subject to forcing corresponding to fig.~\ref{fig:sketch}b.
    Initially, we start from a unit equilateral triangle with uniform conductivities ($k_{ij} = 1$)  In the long-time limit the conductivities stabilize at $k_{ab} = 0.089, k_{bc} = 0.304, k_{ca} = 0.652$ which corresponds to $\delta =k_{ab}/k_{ca} =  0.136$. 
    \textbf{b)} The final values of $\delta$ in simulations ran for different values of the flows feeding a unit equilateral triangle with initially uniform conductivities. Gray area represents all possible stationary configurations. Dark blue dot corresponds to the inflows depicted in Fig.~\ref{fig:sketch}b. Boundary of the allowed region is occupied by forcings in which two out of three $\vec{Q_a},\vec{Q_b},\vec{Q_c}$ vectors have equal magnitude.  \textbf{c)} Stability exponents $\frac{\partial}{\partial C_{i}} (\frac{\text{d}}{\text{d}t} C_{i})$ for a range of values of $f_a$, with other forcings kept constant ($q_a = 0.70$, $q_b = 0.07$, $f_b = 0.35$) give rise to loops only for the intermediate values of fluctuation amplitude $f_a$.}
    \label{fig:trajectory_and_stability}
\end{figure*}

Taking into account that $\text{d}C_{ab}/\text{d}t  = 0$ for $C_{ab} = 0$ we find that the eigenvalue of the Jacobian matrix corresponding to the eigenvector with the nonzero component ${ab}$ is equal to $\frac{\partial}{\partial C_{ab}} (\frac{\text{d}}{\text{d}t} C_{ab})$. Evaluating the derivatives, we find that solution \eqref{eqn:fixed_wall} is stable iff
\begin{equation}
\frac{
        \langle Q_a Q_b \rangle
    }
    {
        \sqrt{\langle Q_a^2 \rangle}\sqrt{\langle Q_b^2 \rangle}
    } 
> \frac{
        l_{bc}^2+l_{ac}^2-l_{ab}^2
    }{2 l_{bc}l_{ac}}. 
\label{eqn:stability_criterion}
\end{equation}
We see that the stability of a tree-type configuration determined by two terms: one capturing the flux fluctuations and capturing the loop geometry (left and right hand sides of Eq.~\eqref{eqn:stability_criterion}). Both sides are dimensionless, which highlights the special scaling properties of the $\gamma = 1/2$ case. Scaling all flows (or lengths) by the same factor does not change the dynamics of the system itself; in that sense, the case $\gamma = 1/2$ is scale-free, and the values of $a$ and $\tau$ have no influence on the typical behaviour of the system.

Our model is fully determined by $\langle Q_i Q_j \rangle$ which can be computed for any reasonable set of forcings. 
The general theory admits an elegant visual interpretation in a slightly less general, but highly relevant, case where driving fluxes are fully correlated. In such cases we can describe each forcing with just the mean $q_i=\langle{Q_i(t)}\rangle$ and mean squared variation $f_i^2$, such that $\langle Q_i(t)Q_j(t) \rangle = f_i f_j + q_i q_j$. Such a correlated behaviour is not uncommon in many transport networks, for example: tidal forces change flows in all the network with an approximately daily pattern, most of the leaf stomata open following the day-night cycle, city traffic often follows a bi-daily pattern etc. The simplest concrete example of such forcing could be $Q_i(t) = q_i + f_i \sqrt{2} \sin(\omega t)$.

When $q_i$ quantifies the constant component and $f_i$ the variable component of the influx $i$ into the system, the left hand side of \eqref{eqn:stability_criterion} corresponds to an angle between $(q_a,f_a)$ and $(q_b,f_b)$  vectors in the $(q,f)$ plane as shown in Fig.~\ref{fig:trajectory_and_stability}. With the vector angle formula and the law of cosines the Eq.~\eqref{eqn:stability_criterion}, the stability criterion for link $ab$, is reduced to
\begin{equation}
    \alpha_{ab} < \beta_{c} \label{eqn:angles_inequality}
\end{equation}
where $\beta_c$ is the angle between the sides $ac$ and $bc$ of the physical network and $\alpha_{ab}$ is the corresponding angle between the flow vectors, see Fig. \ref{fig:sketch}. Eqs.~\eqref{eqn:stability_criterion} and \eqref{eqn:angles_inequality} determine the stability of one link in the network, but terms can easily be rotated to check the stability of each possible tree-like configuration. If none of the inequalities are satisfied, then all possible trees are unstable and the system necessarily has a loop at equilibrium.

Qualitatively, Eq.~\eqref{eqn:angles_inequality} shows that loops should not form whenever the points $Q_a$, $Q_b$, and $Q_c$ in the $(q,f)$ plane are close to being colinear because this ensures that one $\alpha_{ij}$ is small. This happens, for example, when the system is fully controlled by the constant influx. For coastal river networks, this corresponds to the case where tides are minimal, and flows are steady.

Somewhat more surprisingly, the $Q_aQ_bQ_c$ triangle in the $(q,f)$ plane is degenerate also if the system is entirely driven by fluctuations. In the example of river networks, this would correspond to the coastal marsh fed solely by tidal flows. Such systems are indeed loop-free~\cite{Konkol_2022}. 
This is an example of an unexpected symmetry of our model. Since the constraint in Eq.~\eqref{eqn:angles_inequality} depends only on the angle between the discharge vectors (i.e. Fig.~\ref{fig:trajectory_and_stability}b), the situations without fluctuations and without constant component are related by a rotation in the $(q,f)$ plane that preserves the tree stability.

This result adds nuance to those of Kaiser et al.~\cite{Kaiser_2020}, who showed that when a flux applied at all but one nodes are represented by Gaussian independent identically distributed random variables (with the final node satisfying flux conservation), all levels of variance beyond a threshold produced stable loops. However, we show that even large variances will not support a loop if they are highly correlated, as the three discharge vectors will become approximately colinear in the $(q,f)$ plane. Such colinear vectors will satisfy Eq.~\eqref{eqn:angles_inequality} meaning that a tree is stable.

We calculated trajectories with the dynamics given by Eq.~\eqref{eqn:dynamial_system} for an equilateral triangle with uniform initial conductivities ($k_{ij} = 1$) of all sides (Fig.~\ref{fig:trajectory_and_stability}b). The equilibrium configuration is represented by $\delta$: the ratio of the smallest and the largest conductivity in the final network. The gray region in Figure ~\ref{fig:trajectory_and_stability}b shows $\delta$ as a function of the minimum angle between the flow vectors, $\alpha_{min}$. For $\alpha_{min} < 60^{\circ}$ we find that all the final stable configurations are tree-like, i.e. $\delta = 0$ (marked by a horizontal line in Fig.~\ref{fig:trajectory_and_stability}b). This is precisely the angle range in which tree-like solutions should be stable, according to Eq.~\eqref{eqn:angles_inequality}.  In contrast, for $\alpha_{min} > 60^\circ$ all the stable configurations form a loop, with $\delta >0$. This shows that loops and trees never coexist as stationary solutions, which demonstrates that the linear stability analysis fully determines the stability of the system. The dark blue dot corresponds to the flows depicted in Fig.~\ref{fig:sketch}b. In this case, all angles are larger than $60^\circ$, consequently all tree-like configurations are unstable and the final configuration is loopy.

Analysis of stability exponents confirms that loops disappear when the variability is either excessively high or excessively low. For the analysis presented in Fig.~\ref{fig:trajectory_and_stability}c), we fix $Q_a$ as well as the constant component of $Q_b$ while varying the fluctuating component $f_b$, demonstrating that the loops appear only for the intermediate values of the fluctuation amplitude.

We note that although the value of $\alpha_{min}$ determines whether the system will form a loop, it does not determine the exact equilibrium configuration of the loop, because more than one set of $\vec{Q}_i$ exist for a constant $\alpha_{min}$

In summary, we have found an exact condition to maintain any particular loop. Our work is complementary to advances in understanding optimal supply networks \cite{Katifori_2010,Kaiser_2020} and network dynamics \cite{Hu2013} which focus on global properties of the networks rather than individual loops. 

Thus, our results are robust to the change of the number of nodes and individual loop shape. In particular, since $\beta_i$ angles determine the stability boundary we show that even an affine transformation of a network would change its stability properties and thus the choice of equilateral grid by the previous studies on fluctuation-induced loop formation \cite{Kaiser_2020,Katifori_2010,Hu2013} might significantly influence their results.

The loop stability condition \eqref{eqn:angles_inequality} has a particularly simple interpretation when all the driving fluxes are fully correlated, as the stability then depends on the interplay of their constant and fluctuating components, with trees dominating both in small and large fluctuation regimes and loops prevailing in between these two extremes. For more complicated cases, when the fluxes to the nodes are not fully correlated, the stability condition \eqref{eqn:angles_inequality} still holds, but we can no longer interpret the LHS of Eq.~\eqref{eqn:angles_inequality} as an angle in the $(q,f)$ plane. 

Adaptive transport networks can be far more complex than the single loop studied here, both with morphologies that may contain many loops at different scales \cite{passalacqua_geomorphic_2013,ronellenfitsch_topological_2015}, and with complex forcing patterns \cite{bain_flow_2019}. Even so, these networks often adapt toward an equilibrium where conductivities remain relatively constant \cite{wilson_widespread_2017,jarriel_global_2021}. Under the condition that the network is at equilibrium, the presented result should hold, namely that the flows applied at the boundaries of any loop in the system should destabilise trees based on Eq.~\eqref{eqn:stability_criterion}. Hence, even if a network's complexity is such that flow within it cannot be confidently modelled, the network's structure can provide constraints on the scale of fluctuations at the elementary scale.

\begin{acknowledgments}
The work of ML and RW was supported by the National Science Centre of Poland grant Sonata to ML no. 2018/31/D/ST3/02408. PS was supported by the National Science Centre of Poland, grant no. 2022/45/B/ST8/03675. 
JBS was supported by a Fulbright Fellowship US/2021/37/SC and NSF grant EAR-1848993.
We thank Stephane Douady, Adam Konkol and Stanisław Żukowski for helpful and motivating discussions.
\end{acknowledgments}

\bibliography{sources}

\end{document}